\documentclass[12pt,preprint]{aastex}

\def\gtorder{\mathrel{\raise.3ex\hbox{$>$}\mkern-14mu
             \lower0.6ex\hbox{$\sim$}}}
\def\ltorder{\mathrel{\raise.3ex\hbox{$<$}\mkern-14mu
             \lower0.6ex\hbox{$\sim$}}}

\shorttitle{Spitzer Search of DAZs}
\shortauthors{Debes et al.}

\begin{document}
\title{Cool Customers in the Stellar Graveyard IV: Spitzer Search for Mid-IR excesses Around Five DAs\footnote{Based in part on observations made with the NASA/ESA Hubble Space Telescope, obtained at the Space Telescope Science Institute, which is operated by the Association of Universities for Research in Astronomy, Inc., under NASA contract NAS 5-26555. These observations are associated with program \#10560}}
\author{John H. Debes\altaffilmark{1}, Steinn Sigurdsson\altaffilmark{2}, Brad Hansen\altaffilmark{3}}

\altaffiltext{1}{Department of Terrestrial Magnetism, Carnegie Institution of Washington, Washington, D.C. 20015}
\altaffiltext{2}{Department of Astronomy \& Astrophysics, Pennsylvania State
University, University Park, PA 16802}
\altaffiltext{3}{Department of Astronomy, UCLA, Los Angeles, CA 91125}

\begin{abstract}
Hydrogen atmosphere white dwarfs with metal lines, so-called DAZs,
require external accretion of material to explain the presence of weak metal
line absorption in their photospheres.  The source of this material is currently unknown, but could 
come from the interstellar medium, unseen companions, or relic planetesimals
from asteroid belt or Kuiper belt analogues.  Accurate mid-infrared photometry
of these white dwarfs provide additional information to solve the mystery of 
this accretion and to look for evidence of planetary systems that have 
survived post main sequence evolution.  We present {\em Spitzer} IRAC 
photometry accurate to $\sim$3\% for four DAZs and one DA with circumstellar
absorption lines in the UV.  We search for excesses due to unseen companions or
 circumstellar dust disks.  We use {\em Hubble Space Telescope} NICMOS 
imaging of
these white dwarfs to gauge the level of background contamination to our 
targets as well as rule out common proper motion companions to WD 1620-391. 
 All of our targets
show no excesses
due to companions $>$20 M$_{J}$, ruling out all but very low mass companions 
to these white dwarfs at all separations. No excesses due 
to circumstellar disks are observed, and we place limits on what types 
of disks may still be present.  
\end{abstract}   


\section{Introduction}
White dwarfs have long been used to probe the low mass end of the IMF to look
for low mass stellar and brown dwarf companions \citep{probst82,zuckerman92,farihi05}.  With
the advent of more sensitive ground- and space-based imaging at longer wavelengths, the direct detection of substellar objects and planets with a few times
Jupiter's mass around white dwarfs is now possible \citep{ignace01,burleigh02,friedrich05,farihi05,debes05b,debes05a}.

Searching a subset of white dwarfs that harbor markers for 
substellar objects can maximize the return of such a survey.  Nearby hydrogen
white dwarfs with metal line absorption (DAZs) may fit this criterion.  Three
hypotheses have been put forth to explain the presence of DAZs: interstellar
matter (ISM) accretion \citep{dupuis92,dupuis93a,dupuis93b,koester06}, 
unseen companion wind accretion \citep{zuckerman03},
 and accretion of 
volatile poor planetesimals \citep{alcock86,debes02,jura03}.  

ISM accretion
has a wealth of problems in predicting many aspects of DAZs such as the large 
accretion rates required for some objects and the distribution of these objects
with respect to known clouds of dense material \citep{aannestad93,
zuckerman98,zuckerman03,kilic07}.  The quick atmospheric 
settling times of hydrogen atmospheres imply that the white dwarfs are
 in close proximity with accretionary material.

There are roughly 40 cool DAZs known \citep{zuckerman03,koester06}.  Of them,
seven have dM companions, supporting the argument that
 DAZs could have unseen companions that place material onto the 
WD surface through winds \citep{zuckerman03,debes06b}.  In order to accrete enough material, 
 companions must be in extremely
close orbits (P$\ltorder$1 day), bringing into question why these objects have yet to be discovered
through radial velocity surveys of compact objects or
observable excesses in near-IR flux.  In most cases
the reflex motion from such objects would be easily detectable, on the order 
of a few to tens of km/s
\citep{zuckerman92, maxted06}.
  The idea of the presence of unseen companions 
also cannot explain objects like G 29-38 and 4 other white dwarfs which
 have infrared 
excesses due to dust disks within their host white dwarf's 
tidal disruption radius \citep{graham90, patterson91,jura03,zuckerman05,reach05,
kilic06b}.  
The disks around G 29-38 and GD 362 show an amorphous silicate emission feature at
$\sim$10\micron\, implying a small grain size within the disk 
and possibly warped geometries \citep{reach05,jura07}.
Furthermore, companions $>$ 13 M$_{J}$ are ruled out for a wide range of
orbital separations around G 29-38 \citep{debes05a}.

The invocation of cometary or asteroidal material as a method of polluting WD
atmospheres was developed to explain photospheric absorption lines due to metals in
the DAZ WD 0208+395 (G~74-7) \citep{alcock86}.  However, the rates
 predicted by these original studies could not
satisfactorily explain the highest accretion rates inferred for some objects
and could not easily reproduce the distribution of DAZs based on their 
effective temperatures \citep{zuckerman03}. 
However,
mixing length theory predicts a drop-off of observability for
accretion as a function of effective temperature which may swamp out 
the earlier prediction of \citet{alcock86} \citep{althaus98}.  
Also unclear is the 
effect non-axisymmetric mass
loss could have on the fraction of comet clouds lost by their hosts during
post main sequence evolution
\citep{parriott98}.  
By hypothesis, cometary clouds are the result of 
planet formation, so the long term evolution of planetary systems and their
 interaction with these comet clouds needs to be investigated 
\citep{tremaine92}.

 The loss of a star's outer envelope during post main sequence evolution
specifically affects the stability of planetary 
systems, and can rescue the scenario proposed by \citet{alcock86}.
The
Hill stability criterion against close approaches for two comparable mass
 planets qualitatively describes what happens to a planetary system.  
The stability criterion in this case is
 $\Delta_c=(a_1-a_2)/a_1=3\mu^{1/3}$, where
$a$ is the semi-major axis, $\mu$ is the mass ratio of the planets to 
the host star, and $\Delta_c$ represents the critical separation at which
the two planets become unstable to close approaches \citep{hill86,
gladman95}.  During adiabatic mass loss, companions expand their orbits
 in
a homologous way, increasing their orbital semi-major axes
 by a factor M$_i$/M$_f$ \citep{jeans24}. 
The critical
separation grows as the relative separation of the two planets stays the
same, resulting in marginally stable systems being tipped over the edge of 
stability.  This instability can lead to orbital rearrangements,
the ejection of one planet, and collisions \citep{ford01}.
These three events dramatically change the dynamical state
of the planetary system. A fraction of unstable systems will perturb
a surviving Oort cloud or Kuiper belt analogue and send a shower of comets into the inner system
where they tidally disrupt, cause dust disks, and slowly settle onto the
WD surface.  This modification of the comet impact model can explain the
accretion rates needed for the highest abundances of Ca observed and
 the presence of infrared excesses around WDs \citep{debes02}.

The model of \citet{debes02} can be extended to asteroidal material closer to 
the star.  As the central star's mass changes, the basic resonances associated
with any planets will change and bring fresh material into unstable orbits.  
The amount of pollution will depend on the different timescales for comets and
asteroids to be perturbed toward the white dwarf as well as the ratio of 
objects in either asteroidal or cometary orbits.  Asteroids
should be perturbed relatively quickly, on timescales of 10$^8$ yr, while
comets can take up to an order of magnitude longer to be perturbed.  Without 
a more detailed model, however, it is hard to say which population is responsible for DAZ pollution.  

Nine DAZs have already been searched for substellar companions at intermediate orbital separations (10~AU $<$ a $<$ 50-100~AU) with NICMOS high contrast imaging
and AO imaging \citep{kuchner98,debes05b,debes05a,debes06}.  No planets
$>$10 M$_J$ were detected for four, and no brown dwarfs $>$29 M$_J$ were detected for the other five.  Additionally, no unresolved companions were detected
down to substellar limits, following a general finding for a dearth of 
substellar objects around white dwarfs \citep{farihi04,dobbie05,farihi05}.

With the launch of {\em Spitzer} an unprecedented sensitivity is now possible
to further constrain the presence of companions in close orbits, as well as
the presence of dusty disks.  A large interest in infrared excesses
around white dwarfs in general is evidenced by the many surveys of white
dwarfs with {\em Spitzer} \citep{hansen06,kilic06,mullally06,vonhippel07,jura07,jura07b}.

In this paper we present results of our search of four nearby DAZs and 
a DA with circumstellar absorption that have
no known excesses for companions and circumstellar disks.  In \S \ref{sec:IRAC}
we detail our {\em Spitzer} IRAC photometry and results, while in \S \ref{sec:NICMOS} we present second epoch NICMOS images of WD 1620-391 to look for 
common proper motion companions to the white dwarf.  Finally in \S \ref{sec:conc} we present our conclusions.

\section{Spitzer Photometry}
\label{sec:IRAC}
\subsection{Observations}
Table \ref{tab:targs} shows our target DAZs, complete with known T$_{eff}$, log g, distances, and ages.  Cooling ages were taken from the literature and 
initial masses and main sequence lifetimes were calculated by the equations
of \citet{wood92}:

\begin{eqnarray}
\label{eqn:mass}
M_i & = & 10.4\ln{\frac{M_{WD}}{0.49 M_\odot}} \\
t_{MS}& =& 10 M_i(M_\odot)^{-2.5} Gyr.
\end{eqnarray}
Each target was observed with the four IRAC channels, with nominal
wavelengths of $\sim$3.6, 4.5, 5.8, and 8.0~\micron\ \citep{fazio04}.  The observations were
carried out in the mapping mode, with 30 random point dithers for each pair of
channels.  At each dither point, the camera integrated for 100~s,
for a total of 3000~s in each band.  The exception to this was
WD 1620-391, which is a much brighter source.  The images
 had exposure times of 30~s
per dither with 75 dithers for a total integration of 2250~s.  Table 
\ref{tab:obs} summarizes our observations.

In order to obtain {\em Spitzer} IRAC 
photometry with an accuracy of $\sim$3\%, we followed
the prescription laid out in \citet{reach05b}.  
We took the BCD files from the latest {\em Spitzer} pipeline
calibrations for each target (S14.0) and created a final, mosaicked
 image using the
MOPEX package \citep{makovoz05}.  Some caution for point source photometry with
IRAC is warranted.  Post-BCD pipeline calibrated mosaics are not of a high enough fidelity for accurate photometry of stellar point sources.  We routinely found that PBCD images returned photometry systematically 2-4\% higher than when
we used MOPEX.  We performed
overlap correction with a default overlap correction namelist, and mosaicking
with the default namelist given in the IRAC data handbook.  For brighter
point sources,
the outlier rejection schemes of MOPEX can spuriously reject good pixels as
cosmic rays due to photon noise larger than the background variation.  
A typical symptom of this is a coverage map file that shows that
many images were thrown out at the position of the target source.  We experienced good results by choosing an UPPER\_ and LOWER\_THRESHOLD parameter of 15 for the MOSAICIN module, as well as using the keyword REFINE\_OUTLIER to ensure bright
sources were treated with a threshold closer to 20.  The thresholds refer to
the number of sigma above the mean background.  As a final check we visually inspected the resulting coverage maps to ensure that most images were used by
the mosaicking program.

Since each of our images had several dither
positions, we did not make any array-location or pixel phase corrections.  We estimate that these effects are at the level of 1\% and not a significant 
error source, but we include them in our total error.  We
performed aperture photometry with a 3 pixel radius ($\sim$3.6\arcsec), and
used a 4-pixel wide annulus starting just outside the source aperture for
background subtraction, to ensure as accurate estimate of the background as possible.
Aperture corrections appropriate for this size
source radius and background annulus were applied, as well as calibration
factors,  flux conversions and a 
color correction in each band assuming a $\nu^2$ spectral slope as mentioned in \citet{reach05b}.   
The consistency of both aperture corrections and the photometry with different
sized apertures was checked by recalculating the photometry with 5 pixel
radius apertures with background annuli with 5-pixel radii starting just outside the source aperture, and 3-pixel source apertures with 10-pixel wide annuli starting at a radius of 10 pixels.  We avoided a 2-pixel source aperture as that
appeared to consistently give photometry lower by $\sim$2-5\%.  For channels 1 and 2, differences between the three choices were 
never more than 1\% except in the case of WD 0245+541, which has several nearby sources within 4-10 pixels.  Channels 3 and 4 often had larger changes for
the 5 pixel radius aperture, up to $\sim$10\% but typically closer to 2\%.  
We attribute these systematic 
changes primarily to residual structure in the background and to 
coincident sources.  Both of these sources of systematic error are lessened
by the small aperture and small background annulus.  We estimate that on
average there is a 1\% error from sytematic uncertainties in aperture photometry based on our specific choice of aperture and background annulus.
   
\begin{figure}
\plottwo{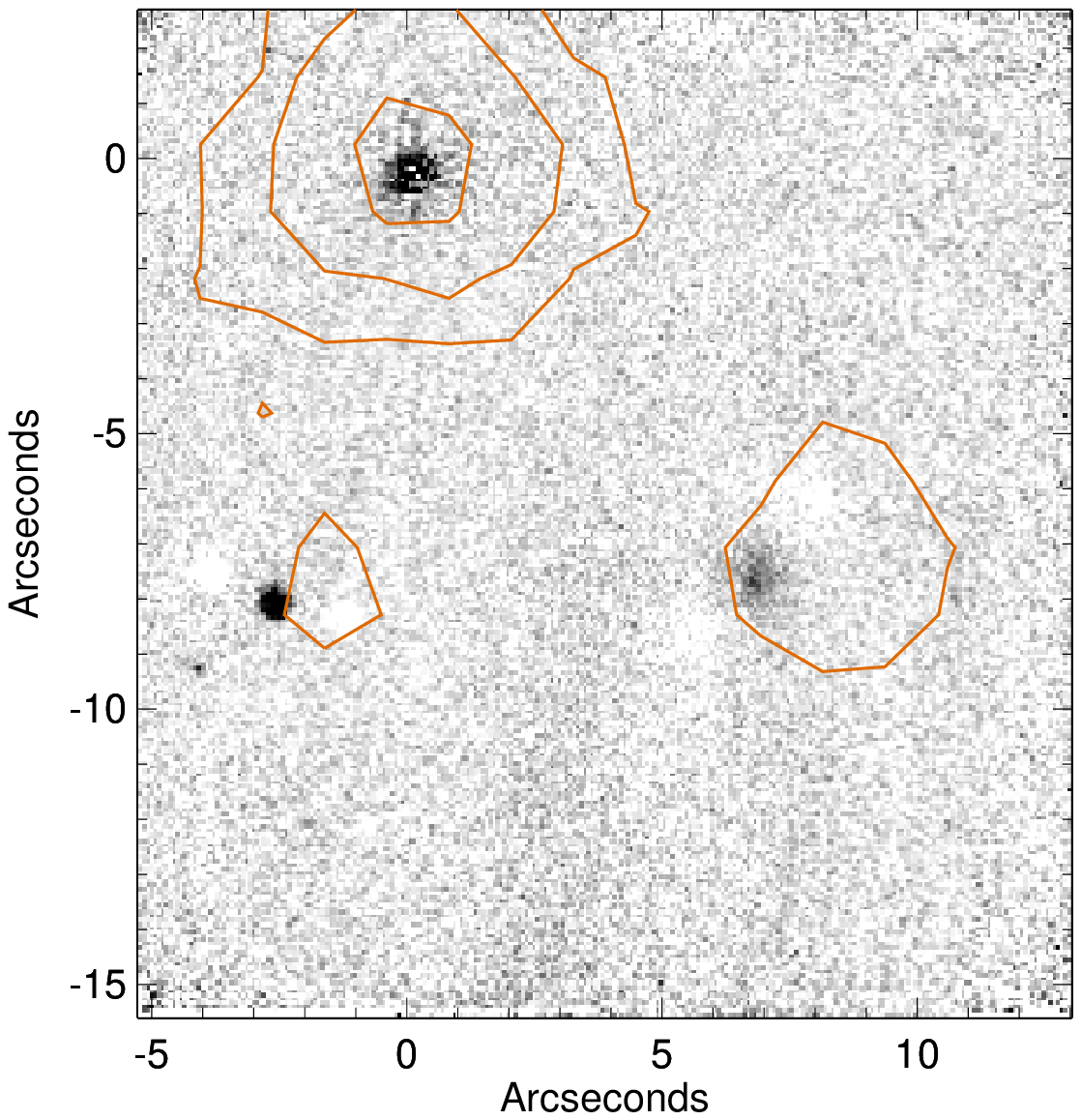}{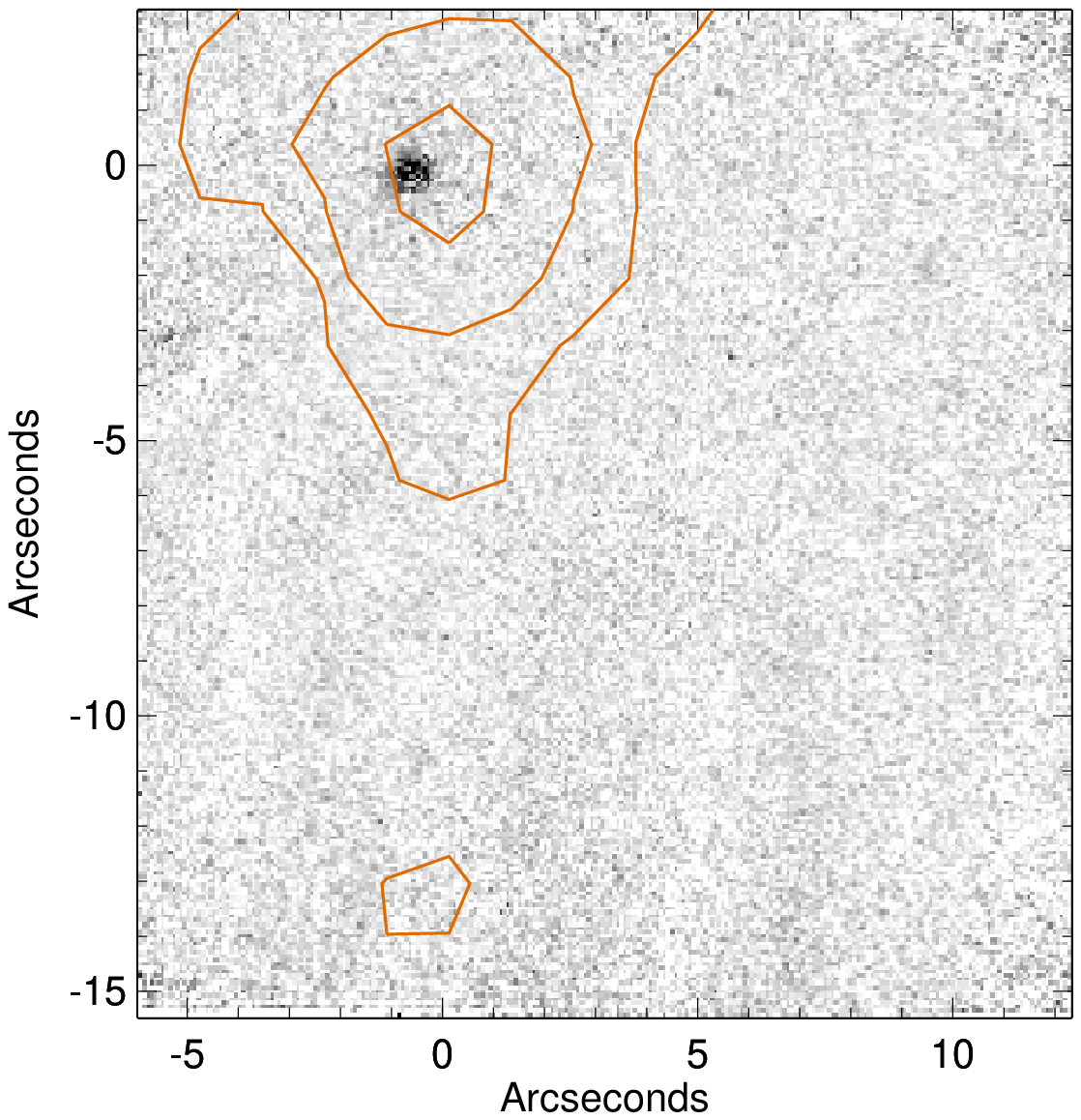}
\caption{\label{fig:wd02} NICMOS F110W images of 
WD 0208+396 (left) and WD 0243-026 (right).  The contours are from IRAC channel
two images where the levels correspond to 0.1\%, 1\%, and 10\% of the total
observed flux from the white dwarf.}
 \end{figure}

\begin{figure}
\plottwo{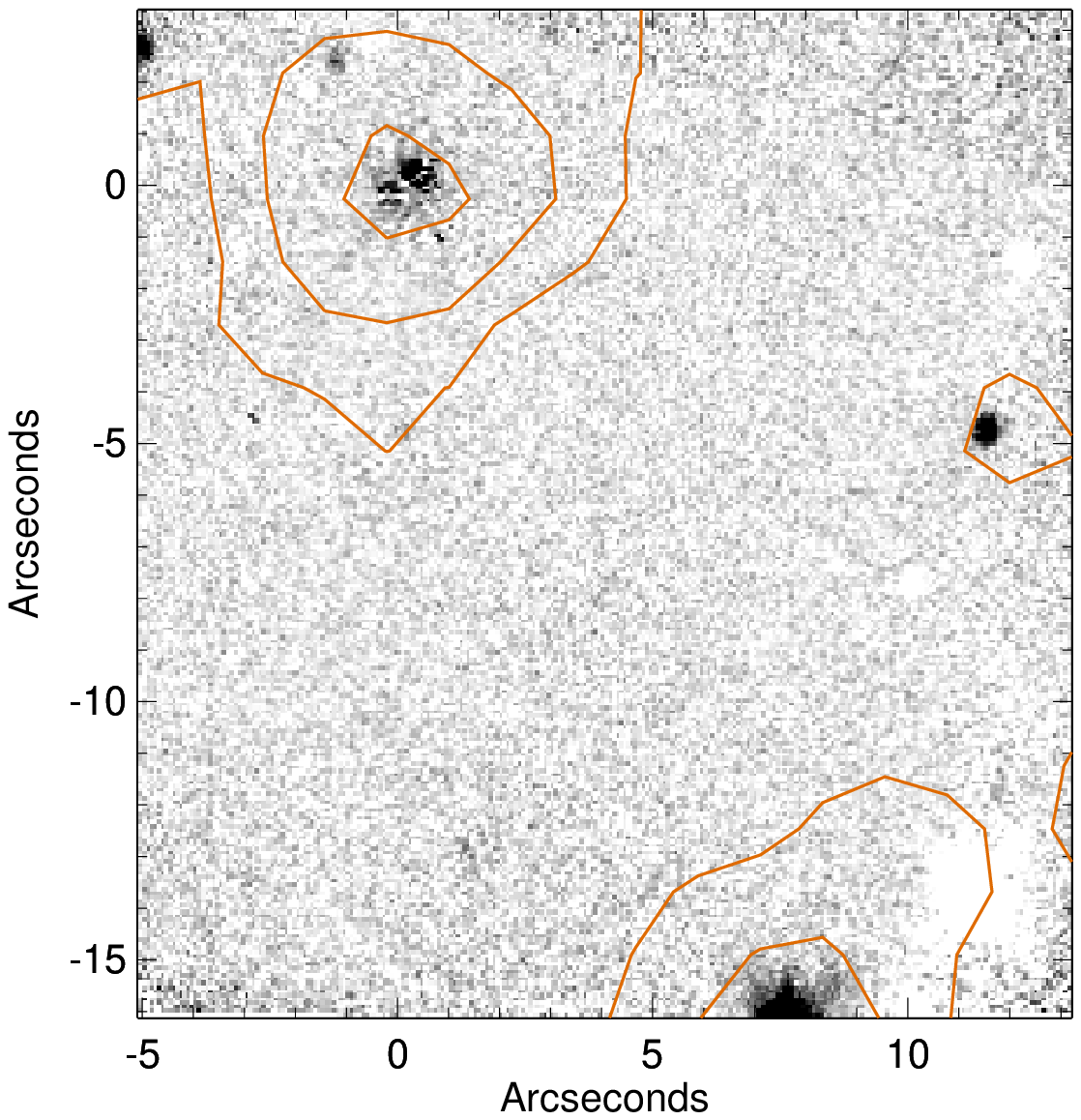}{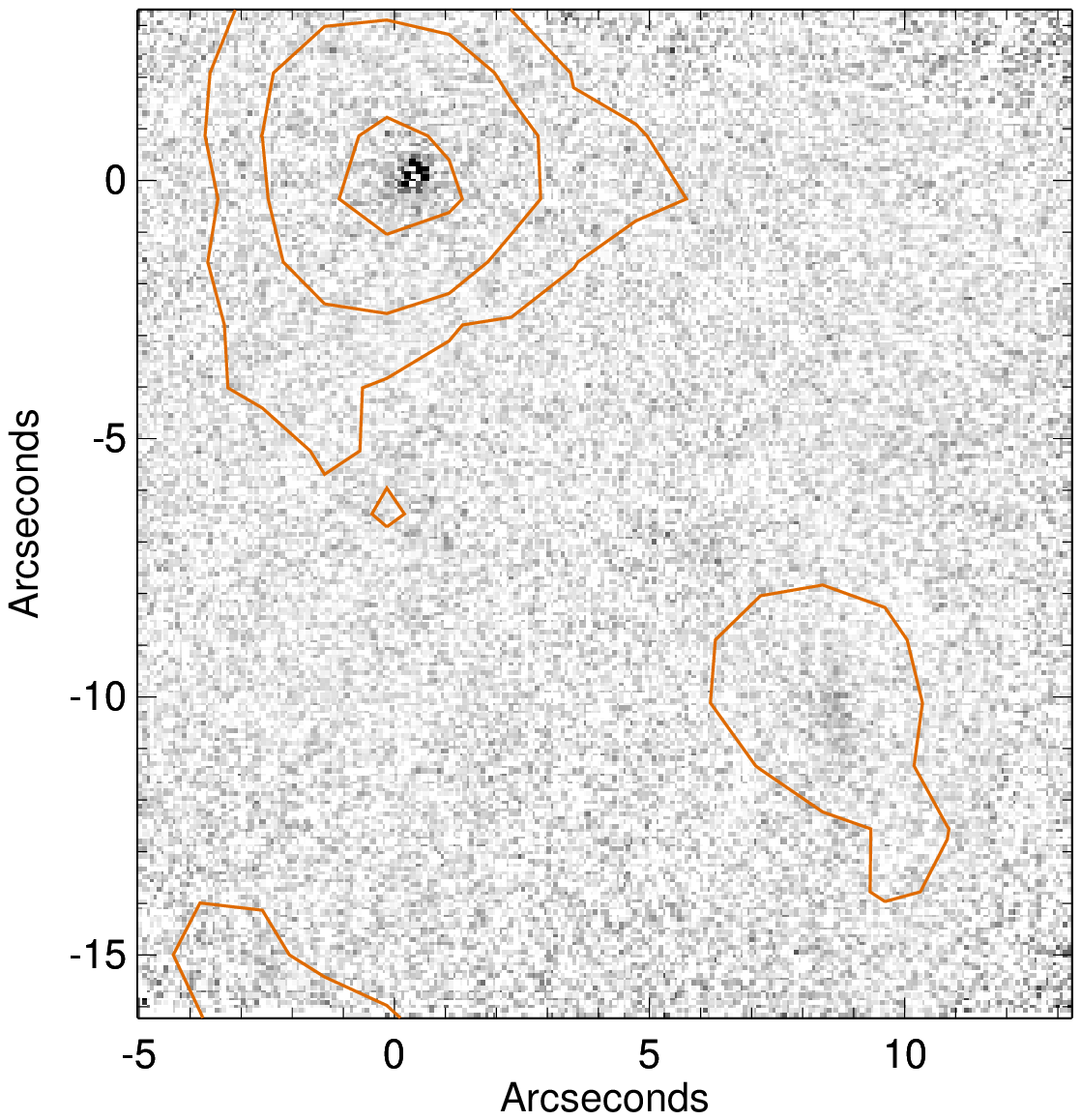}
\caption{\label{fig:wd541} NICMOS F110W images of 
WD 0245+541 (left) and WD 1257+271 (right).  The contours are from IRAC channel
two images where the levels correspond to 0.1\%, 1\%, and 10\% of the total
observed flux from the white dwarf.}
\end{figure}

\begin{figure}
\plotone{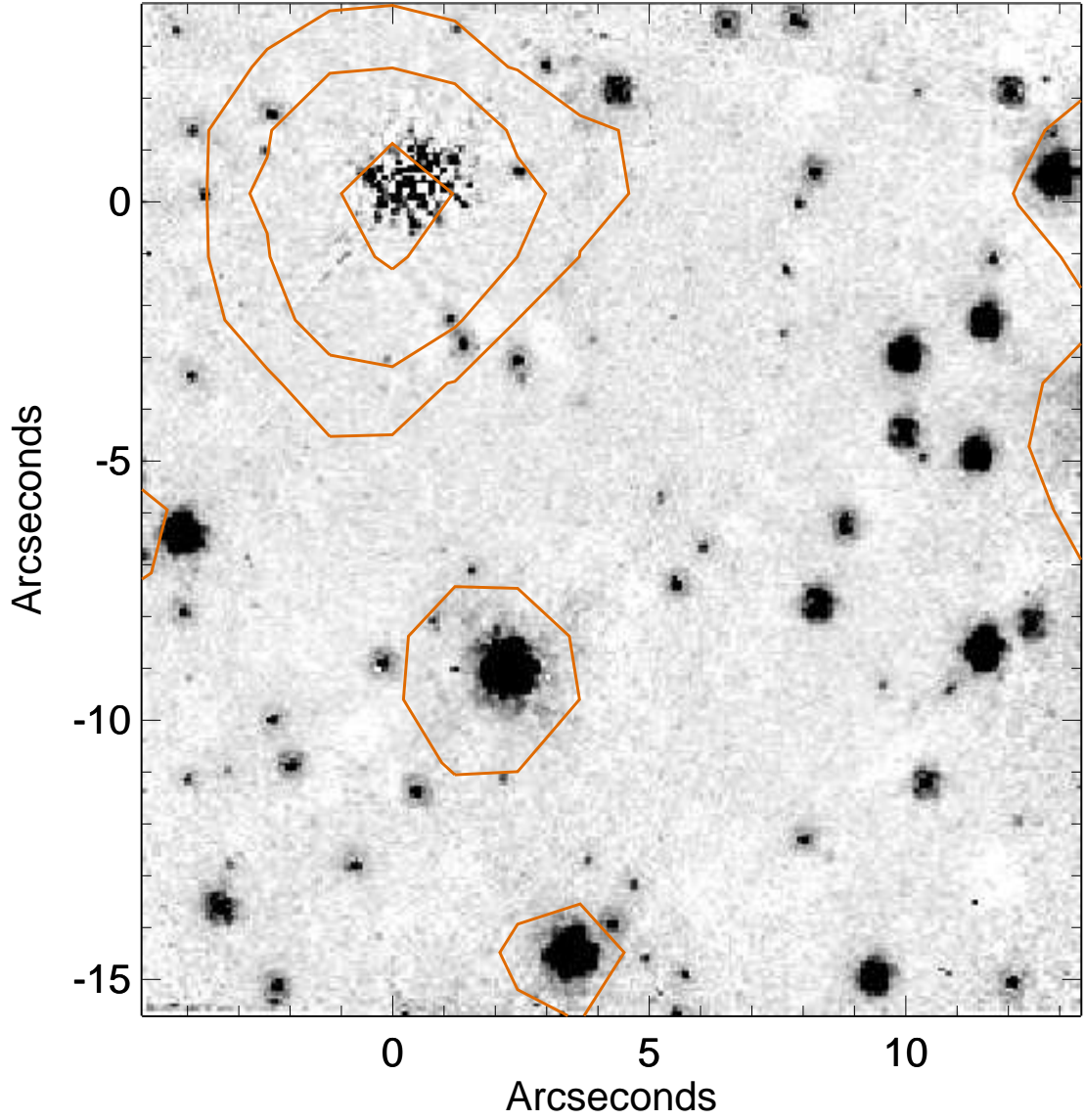}
\caption{\label{fig:wd16} NICMOS F160W image of 
WD 1620-391.  The contours are from IRAC channel
two images where the levels correspond to 0.1\%, 1\%, and 10\% of the total
observed flux from the white dwarf.}
\end{figure}

No obvious interstellar cirrus was noted
for any of our targets in the 8\micron\ channel.  Figures \ref{fig:wd02}-
\ref{fig:wd16} show PSF subtracted NICMOS images
of the DAZs from \citet{debes05b}, with contours from the final IRAC channel
2 images
overlaid.  The contour lines correspond to 0.1\%, 1\%, and 10\%
of the total measured flux 
respectively to demonstrate the absence of contaminating 
objects in the source and background photometric apertures.  The target WD in 
each image is located at the point (0,0), and appears as a speckled area 
since it is behind the coronagraph and the residual PSF has been subtracted off.

For the observations of WD 0208+396, the IRAC detector was struck by a large
number of solar protons, degrading the images with cosmic ray hits.  
The looser constraints on outlier rejection can give higher counts at
the level of 10\%.  These hits were worse for the 5.8\micron\ channel but
 we used a more stringent threshold for the MOPEX outlier 
routines of 3 for channels 3 and 4 instead of 15.  
Inspection of the coverage maps for the channels
show that most of the images could still be used, with the most images being
rejected for the 5.8\micron\ channel.  We verified that we got consistent 
photometry by visually inspecting individual BCD images and combining only
the files without obvious cosmic ray strikes.

The estimated photometric errors for each channel are quite small due to the large S/N achieved.  In addition to the standard errors in photometry, we added
a 3.3$\%$ factor to account for the overall uncertainty in the flux calibrations
quoted by \citet{reach05b} as well as the contributions from uncertainties mentioned above.

\subsection{Comparison of Photometry to WD models}
In order to detect a bona fide excess, one must compare the observed flux with
an expected flux.  We compared our observations with models
of \citet{bergeron95} as well as the $BVRIJHK$ photometry of \citet{bergeron01}
for four of the five targets.  WD 1620-391 was not part of \citet{bergeron01}'s
survey and so we used a combination of USNOB, Hipparcos, and 2MASS photometry. Fluxes in the mid-infrared were kindly provided (P.E. Tremblay,private communication), using updated models
from \citet{tremblay06} and without any knowledge of the measured mid-IR fluxes.  We further normalized these flux densities to a median of the visible and Near-IR flux densities to account for any slight offsets between the observed
data and the models.  This approach differs from previous work reported, where 
blackbody extrapolations of the WDs' K flux density 
were compared with our {\em Spitzer} data \citep{debes07}.    

For the level of photometric accuracy we have achieved, white dwarfs with 
effective temperature of $\sim$5000-7000~K depart from true black bodies,
mainly due to H$^{-}$ bound-free and free-free opacity, with the free-free
opacity being most important for the near- and mid-infrared (P.E. Tremblay,
private communication).  Free-free absorption can be calculated precisely at
long wavelengths and is incorporated in WD models \citep[see][for example]
{john88}

\begin{figure}
\plotone{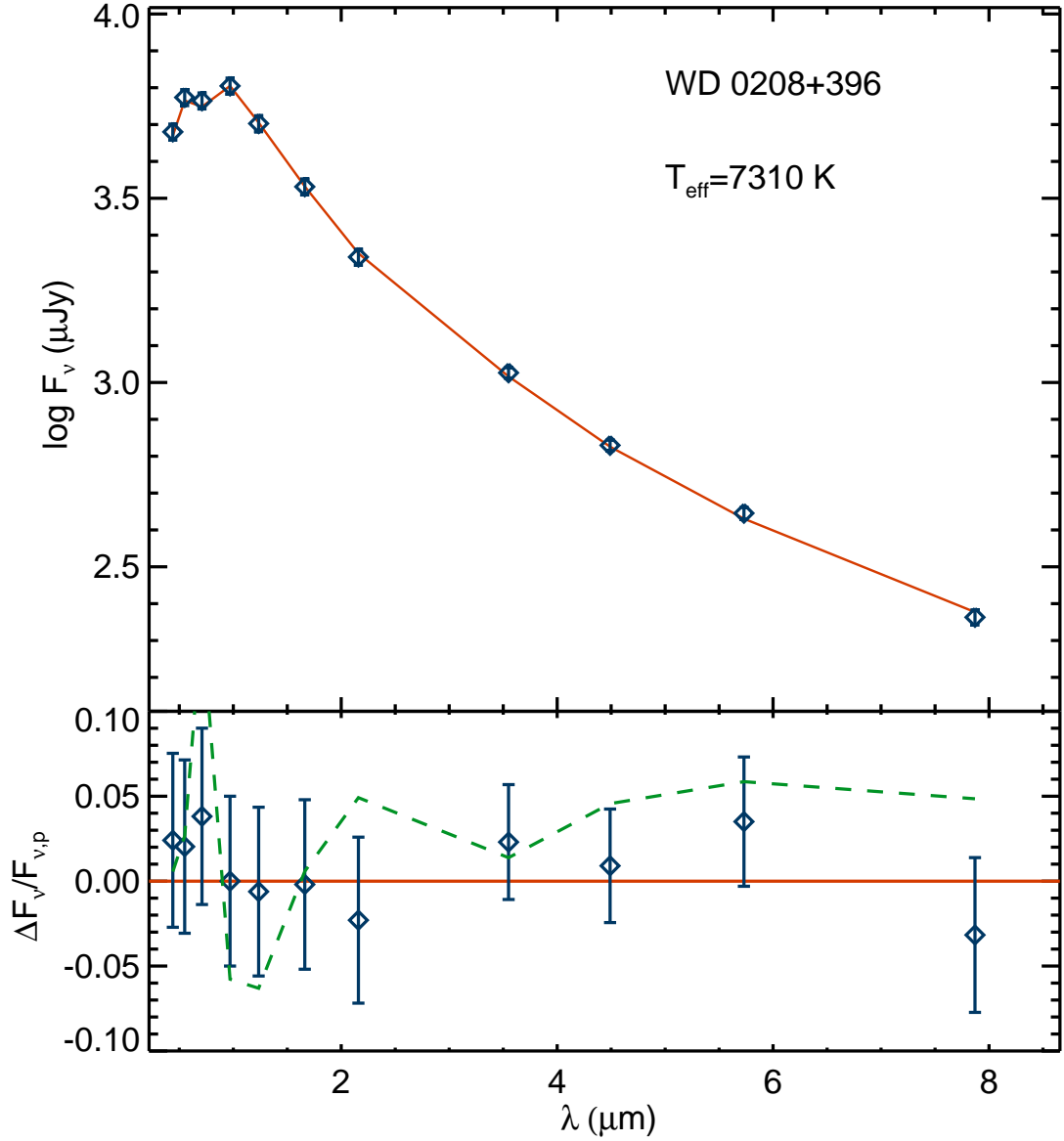}
\caption{\label{fig:74graph} Comparison of observed fluxes for WD 0208+396
(diamonds) and predicted fluxes (solid line) based on the models of \citet{bergeron01}.  The bottom panel shows a close-up of the residuals in the IRAC channels as well as the differences compared to a pure blackbody SED (dashed line).}
\end{figure}

\begin{figure}
\plottwo{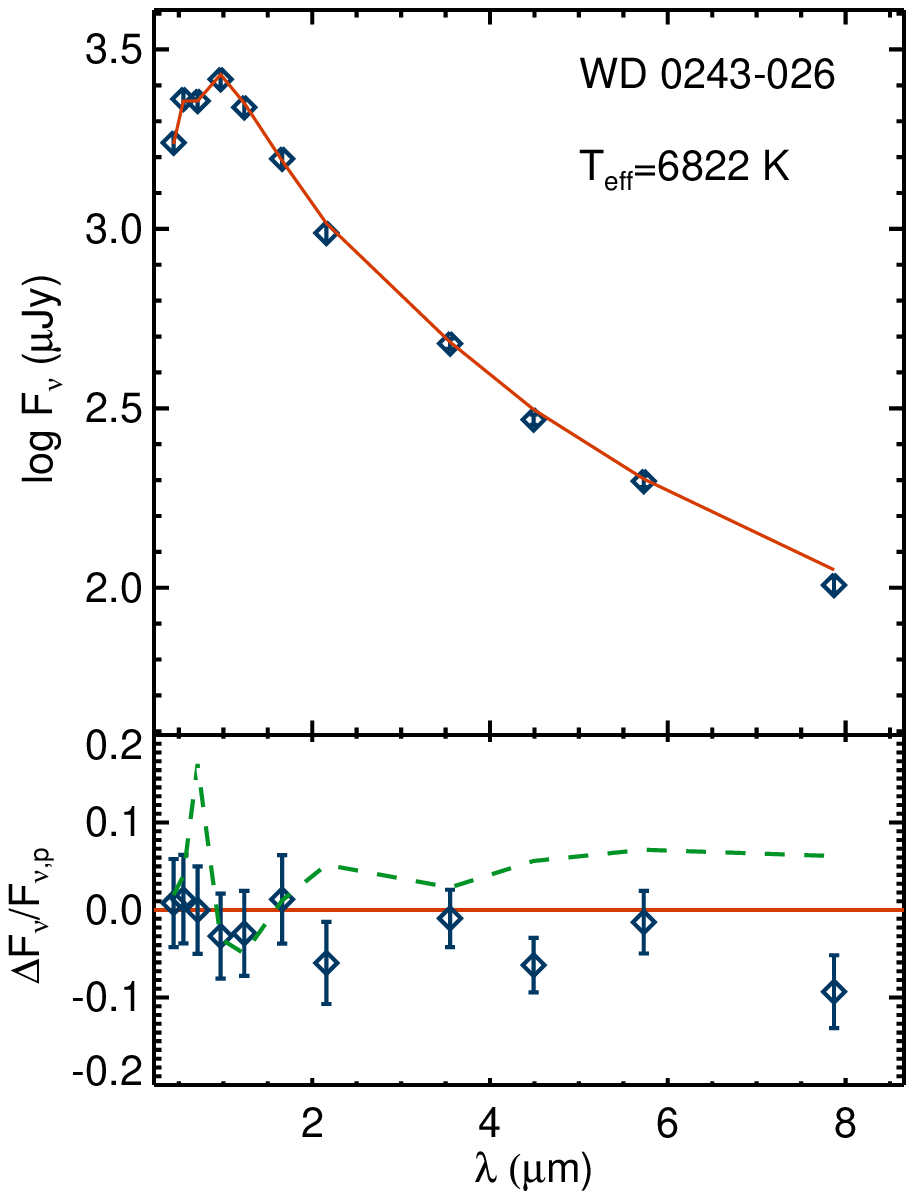}{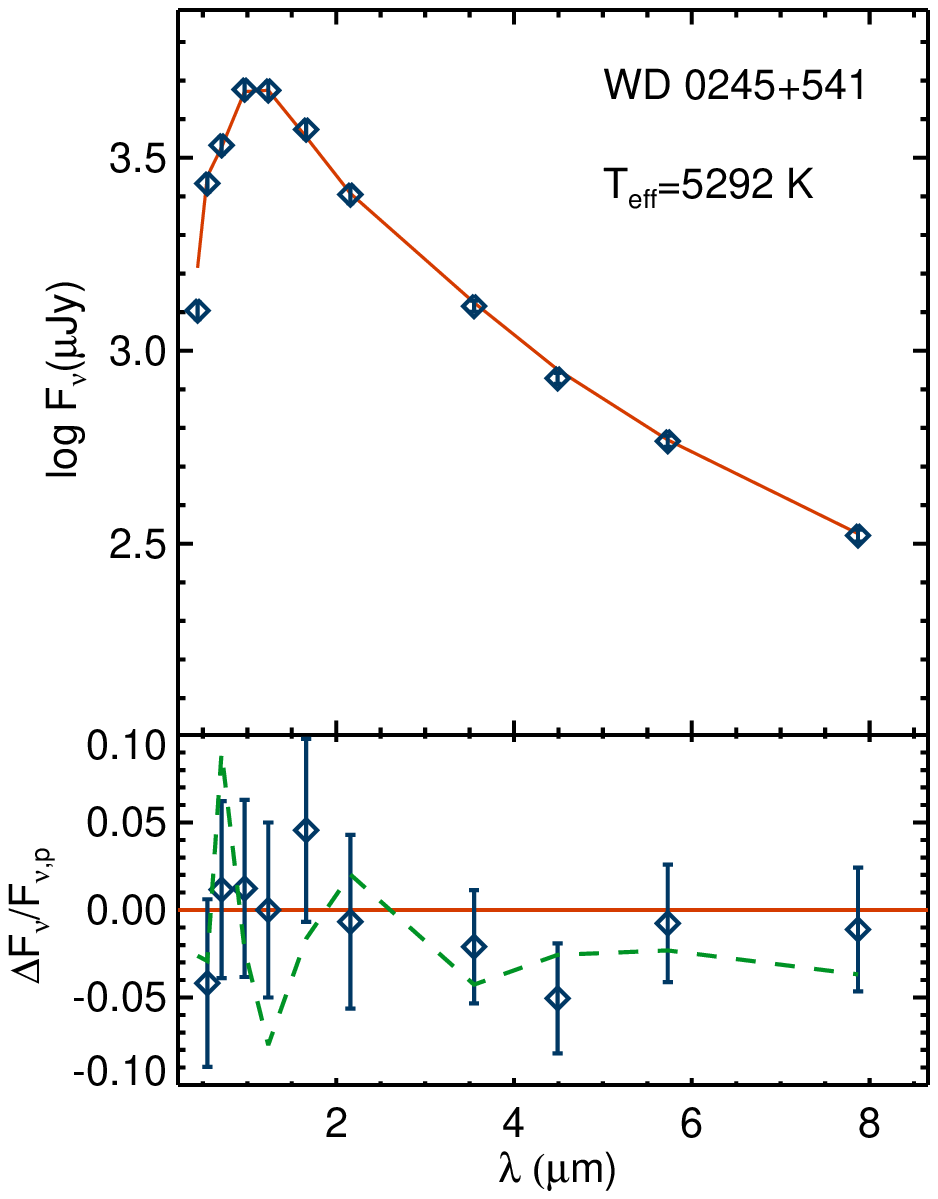}
\caption{\label{fig:panel1} Same as \ref{fig:74graph}, but for WD 0243-026 (left) and WD 0245+541 (right).}
\end{figure}

\begin{figure}
\plottwo{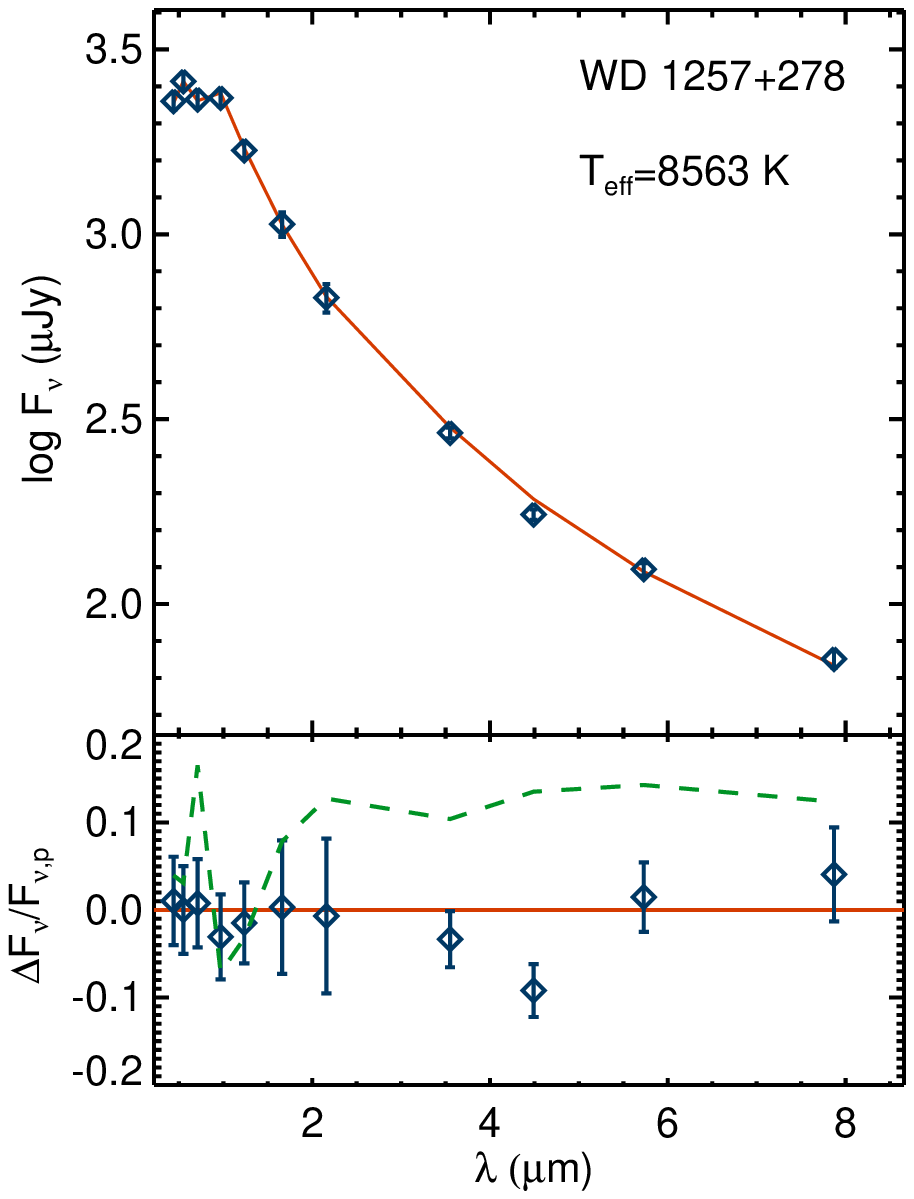}{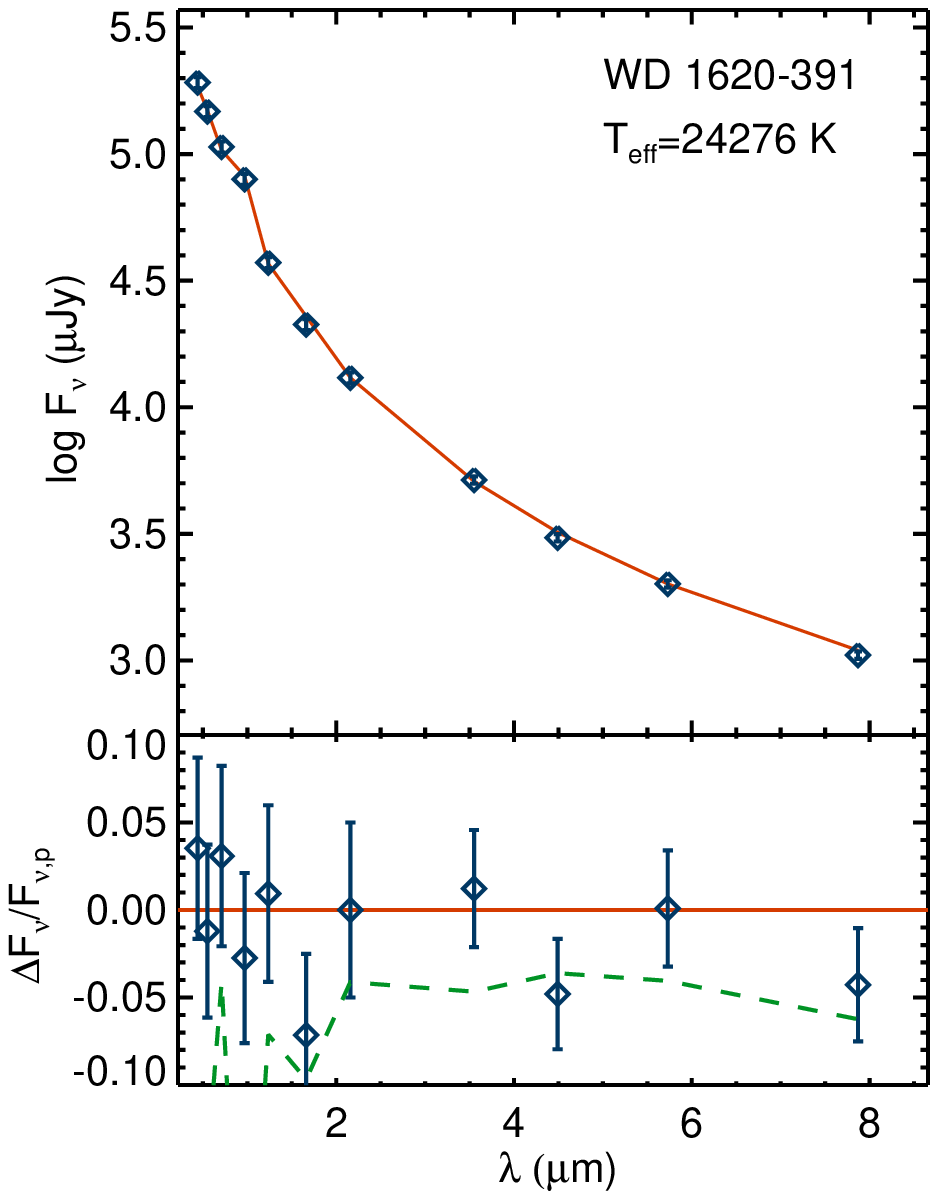}
\caption{\label{fig:panel2} Same as \ref{fig:74graph}, but for WD 1257+278 (left) and WD 1620-391 (right).}
\end{figure}

Figure \ref{fig:74graph} shows a representative comparison between 
the model fluxes and the
measured fluxes for WD~0208+396, as well as the residuals.  
The full list of predicted and observed IRAC 
fluxes for
all of our targets is in Table \ref{tab:fluxes}, while Figures \ref{fig:panel1}
and \ref{fig:panel2} show the SEDs of the remaining targets.  We required that a significant excess (deficit) be
$>$ three times the photometric error above (below) the
calculated model flux in at least one channel.  We find that for the exception
of WD 1257+278, the model fluxes and photometry agree to within 1-2 $\sigma$.

Figure \ref{fig:panel2} shows the SED of WD 1257+278 compared to the model.  There isexactly a 3$\sigma$ deficit in the 4.5 band, to a depth of 10\%.  The mosaic coverage maps show no images being thrown out where
the photometric aperture is located.  A slight mismatch between the model effective temperature and the true effective temperatrue 
could present an artificial deficit or excess, but the errors in the derived
effective temperature are on the order of $\sim$2\%, which would correspond to
errors in the predicted fluxes of 3-4\%, much less than the observed deficit
(P.E. Tremblay, private communication).
Despite matching our criteria for selection as a significant deficit, we believe it is tentative at best, based on a detailed analysis of the match between
our photometry and the models.

Because of the deficit with WD 1257+278 we wished to get
an empirical sense of how well the data matched the predicted model fluxes.
To that end, we took the standard deviation of $\Delta F_\nu/F_{\nu,p}$ in all
the channels where $\Delta F_\nu$ is the difference between the observed
flux density and the predicted flux density ($F_{\nu,p}$), as well as the mean
$\Delta F_\nu/F_{\nu,p}$ for each channel.  We find that the standard deviation of the sample is $\sim$3.7\%, while the mean for each channel is -1\%,-5\%,0.09\%, and -3\%.  These results indicate that the predicted
fluxes match the observed fluxes to within the absolute calibration errors we
assume.  We note that the 4.5\micron\ channel appears to have a barely marginal
($\sim$1.4$\sigma$) mean deficit, with four of the five targets possessing $\sim$
5\% or greater
deficits, WD 1257+278 being one of these objects.  WD 0208-396 is the only object with no deficit at 4.5\micron.  

\begin{figure}
\plotone{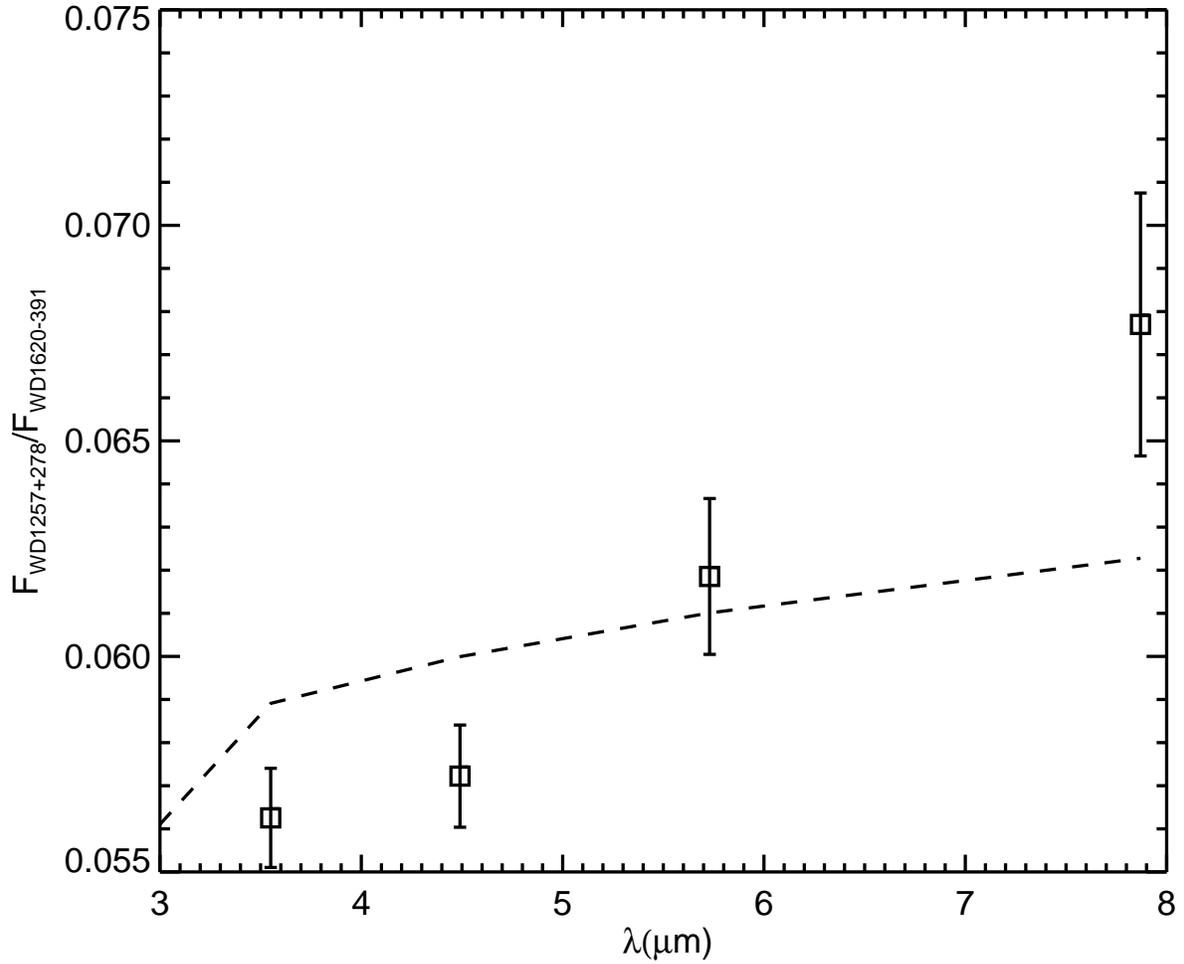}
\caption{\label{fig:comparison} Comparison between the measured flux ratio
of WD 1257+278 to WD 1620-391 (squares) and that predicted by white dwarf models (dashed line).  WD 1257+278 shows a significant deficit in its absolute photometry which is not reproduced relative to WD 1620-391.}
\end{figure}

As another test, we divided the IRAC photometry of our target DAZs by WD 1620-391, the brightest WD in our sample with the highest signal-to-noise.  In this case, we are limited by photon noise and the stability of the IRAC detectors,
which is on the level of $\sim$2\%.  We compared the relative photometry
of WD 1620-391 and WD 1257+278 to the model fluxes in Figure \ref{fig:comparison}.  Within the estimated errors, the observed flux ratios match the expected
ratios.  We repeated this test with the other white dwarfs and found similar
agreement.  The consistency of the flux ratios suggests that the depression of
flux at 4.5 micron may be due to a systematic error in the aperture correction,
color correction, or calibration factors for that channel.

Observed deficits for a white dwarf
may be evidence for circumstellar material raining
down on its surface.  If such a situation were confirmed at 4.5 or 8\micron,
we predict that non-LTE absorption by SiO gas may be present, with
possibly some contribution from CO.
Absorption due to fundamental and overtone rotational-vibrational bands of SiO and CO in late type stars is well known  \cite{cohen}.
The dissociation temperature of SiO and CO are high enough that these species
 could persist at the temperatures of cooler white dwarfs. 

The absorption could be boosted if SiO is formed above the white dwarf photosphere through photodissociation of SiO$_2$ (and any CO present is similarly
formed through photodissociation of carbonates) from refractory dust which sublimates as it is brought down to the the white dwarf surface
through photon drag.
The resulting SiO is formed at low densities just above the photosphere, and is far from local thermodynamic equilibrium, with much larger absorption
strengths than inferred from photospheric LTE.  This absorption would show up
most strongly around 4-5\micron\ and $\sim$8-10\micron\, where SiO has
fundamental and first overtone bands at 8.0 and 4.1\micron, respectively.
CO would show up primarily in the second channel with its fundamental band
at 4.7\micron \citep{cohen}.  The details of this scenario need to be studied
further to determine the feasibility of observing absorption due to SiO or CO
gas.

\subsection{Limits to Companions}
For IRAC, very cool substellar objects can be detected as excesses, especially
due to a ``bump'' of flux for brown dwarfs and planets at $\sim$4.5\micron.
While theoretical models predict the 4.5\micron\ flux to be large, observations
of cool brown dwarfs suggest that the spectral models overestimate this flux
by a factor of $\sim$2 \citep{golimowski04,patten06}.

In order to place upper limits on the types of unresolved
companions present around our targets, we compared predicted IRAC fluxes for
cool brown dwarfs  and planets
by convolving the IRAC filters with the models of \citet{bsl03} appropriate for the particular age of each target DAZ and its distance.  
For the 4.5\micron\ channel we assumed that the resultant flux was a
factor of two smaller than predicted.  We then compared our 4.5\micron\ 
3$\sigma$ limits to those models in order to determine a mass limit.  These
results are presented in Table \ref{tab:limits}.  In all cases we improve the 
unresolved companion limits to these objects over \citet{debes05b} by a factor
of 2-4.  For WD 0243-026 and WD 1620-391 we rule all companions $>$14 M$_J$
 objects for separations $<$76 and 46~AU respectively.

\subsection{Limits to Dusty Disks}

We can determine limits to two types of dusty disks, either geometrically
flat, optically thick disks, such as that modeled for G~29-38 or GD~362,
or diffuse, optically thin disks.  Both GD~362 and G~29-38 can be well 
modeled by disks not unlike Saturn's rings, within the tidal radius of the
white dwarf with an interior edge at the dust sublimation radius \citep{jura03,zuckerman05,jura07,vonhippel07}.

\subsubsection{Optically Thick Disks}
If we assume an optically thick disk, the emission of the grains can be
modeled following \citet{adams87}:
\begin{equation}
\label{eqn:fnu}
F_{\nu}=\frac{2\pi\cos(i)}{d^2}\int^{R_{out}}_{R_{in}} B_{\nu}(T)r dr 
\end{equation}

with T as a function of R:

\begin{equation}
T=\left(\frac{2}{3\pi}\right)^{\frac{1}{4}}\left(\frac{R_\star}{r}\right)^{\frac{3}{4}}T_\star
\end{equation}

This assumes that the inner radius corresponds to a dust 
sublimation radius of 1200~K.  In Table \ref{tab:limits}, we show the upper limits to $i$ based on our
lack of 3$\sigma$ detections in our 8.0\micron\ channel data.  In most cases, excess
emission would have been significantly detected at shorter wavelengths as well.
If this type of disk is present around these DAZs, the inner edge of the
disks must be at $\gtorder0.4$ R$_{\odot}$, or all of them are close to edge-on.  We can quantify the probability of observing 5 systems with inclinations
determined by our upper limits out of a random sample of disk inclinations.  For any one disk, this is $\sim$1-$\cos{i}$, and for all five targets the probability is negligible.  Most optically thick dust disks observed seem to have
exterior radii of $<$0.6 R$_\odot$ \citep{vonhippel07}.

Given the $10^3$-$10^4$ year settling timescales ($t_D$, See Table \ref{tab:targs})
for our targets, the lack of a disk does not necessarily imply that
the DAZs cannot accrete material in this manner.  As \citet{hansen06} has
pointed out, the timescale for removal
of dust grains within the tidal disruption radius of a white dwarf due 
to Poynting-Robertson drag is short:

\begin{equation}
\label{eq:tpr}
T_{PR} = \left(\frac{s}{1\mu \mbox{m}}\right)\left(\frac{\rho_s}{3 \mbox{g cm}^{-3}}
\right)\left(\frac{r}{10^{10} \mbox{cm}}\right)^2 \left(\frac{L_\star}{10^{-3} L_\odot}\right) \mbox{yr}
\end{equation}

where $s$ and $\rho_s$ are the average grain size and density respectively,
and
 $r$ is the distance from the star, ranging from $\sim10^{10}-10^{11}$ cm.
If an incoming comet or asteroid
 is disrupted and all of the material is removed before 
another arrives, then some fraction of the time a DAZ will have this type of
 disk and
at other times it won't while still retaining a detectable
metal line signature.  The metal line will remain detectable as long as the 
metal settling time is roughly longer than the time to the next replenishing
collision.
Cooler dust from collisions may still
be detectable at longer wavelengths, or slowly drift inwards from further away.  Using Equation \ref{eq:tpr}, one can determine the rough orbital separation
from which dust would spiral in over 1 Gyr, or a typical cooling time 
for a white dwarf.  Assuming the typical values in Equation \ref{eq:tpr}, dust could spiral in from as far as $\sim$20~AU.

\subsubsection{Optically Thin Disks}
If we expect an optically thin disk, we see the emission from every emitter.
If one assumes a particular size (and therefore a particular mass) per emitter
and the number of emitters per unit area, one can determine the total mass
in an optically thin dust disk based on the observed flux.
  We focus in particular on the limit to dust
between the tidal radius of the white dwarf and the dust sublimation radius,
since this region is of most interest for explaining DAZ metal accretion.

For the sake of simplicity, we assume that a constant number density of 1~\micron\ dust particles reside in a flat optically thin 
disk between the dust sublimation radius
$R_{sub}$ and the approximate tidal disruption radius, $R_{tidal}\sim\left(\bar{\rho}_{WD}/\rho_{obj}\right)^{\frac{1}{3}}R_\star$ of the DAZ, assuming a $\rho_{obj}$=3 g~cm$^{-3}$ for the parent bodies to the dust.  In this case the flux is given by
a modification of Equation \ref{eqn:fnu}:
\begin{equation}
F_{\nu}=\frac{2\pi s^2\cos(i)}{d^2}\int^{R_{tidal}}_{R_{sub}} n(r)B_{\nu}(T)r dr
\end{equation}
where we have utilized the models of \citet{laor93} to calculate the spherical
1\micron\ 
grain temperature of each dust particle given each DAZs luminosity \citep{bergeron01,bragaglia95}.  For each WD we normalize $n(r)$ such
that the resultant dust disk spectrum returns the 3$\sigma$ flux limit when
convolved with the IRAC 8 \micron\ channel filter response.  Table \ref{tab:limits}
shows the resulting upper limits for dust disk mass.  For WD~1620-391, its radius at which dust sublimates exceeds the tidal disruption radius, 
and so we expect no dust to be 
present in this region.  Similarly hot white dwarfs would not have dusty disks
around them like G~29-38 or GD~362.  They may have gaseous disks around them,
as evidenced by the discovery of a gaseous, metal-rich disk around a hot DA
white dwarf \citep{gansicke06}.
  
If there are dust disks,
then dust accretion could conceivably occur for longer then the DAZ atmospheric
settling times in our sample.  However, the PR drag timescale at the tidal
disruption radius for each DAZ is $\ltorder$ $M_{disk}/\dot{M}$.  This implies
that accretion is not driven by PR drag of a present disk.

\section{NICMOS imaging}
\label{sec:NICMOS}

NICMOS coronagraphic images of these five white dwarfs were presented in \citet{debes05b}, 
with accompanying limits to companions at 1.1\micron, as well as 1.6\micron\
for WD 1620-391.  High spatial resolution NIR images are particularly useful
for discriminating against potential sources of background contamination which
could bias the mid-IR photometry to spurious excesses,
given the IRAC camera's spatial resolution of 1.2\arcsec/pixel.  While it may be 
rare to find coincident sources that may contaminate the photometry of the target, two of the five targets have visual companions within 4\arcsec\ of the
target star.

WD 1620-391, one of the targets with a large number of visual companions,
is close to the galactic plane.  This interesting object is not technically a 
DAZ.  It a DA with no optical metal absorption lines that is a large separation common proper motion companion to a planet bearing star \citep{mayor04}.   In the UV it possesses metallic circumstellar absorption lines \citep{holberg95,wolff01}.  The planet bearing star is separated by 5\farcm 75 (4451~AU), and is well off the field-of-view for NICMOS.  
Even expecting a large number of coincident
sources due to its galactic latitude, it possessed an overdensity over that 
expected
\citep{debes05b}. 
Motivated by this overdensity, a second epoch image of 
WD 1620-391 was obtained in March 2006, two years after the first image was
taken to search for any common proper motion companions.  The new image was
reduced following the basic prescription laid out in \citet{debes05b}, 
where the white dwarf was imaged at two separate spacecraft roll orientations
and each roll image was subtracted from the other and combined to produce a 
high contrast final image.  The other objects in the field were masked out in the
image that was used as a PSF reference, since the field of view was moderately
crowded and subtraction residuals would hamper the detection of faint sources.

We aligned both epochs on the pixel position of WD~1620-391 and rotated the
images so that North was in the positive vertical direction of the images,
using pixel centers and orientations as header keywords from the STScI pipeline.  We then shifted the second epoch image
by the measured proper motion of WD 1620-391 of 97.49$\pm$3.28 mas/yr ($\mu\cos{\delta}$=75.52 mas/yr) in right ascension and 0.05$\pm$1.74 mas/yr in declination \citep{hip} to align the shifted background stars.  We measured the centroids
of $\sim$70 observed objects common in both fields using the IDL ASTROLIB routine GCNTRD and measured the difference in centroid position from one epoch to
another.  With this procedure, any object co-moving with WD 1620-391 would
have a position shift of 2.58 NICMOS pixels, or 0\farcs19.

\begin{figure}
\plotone{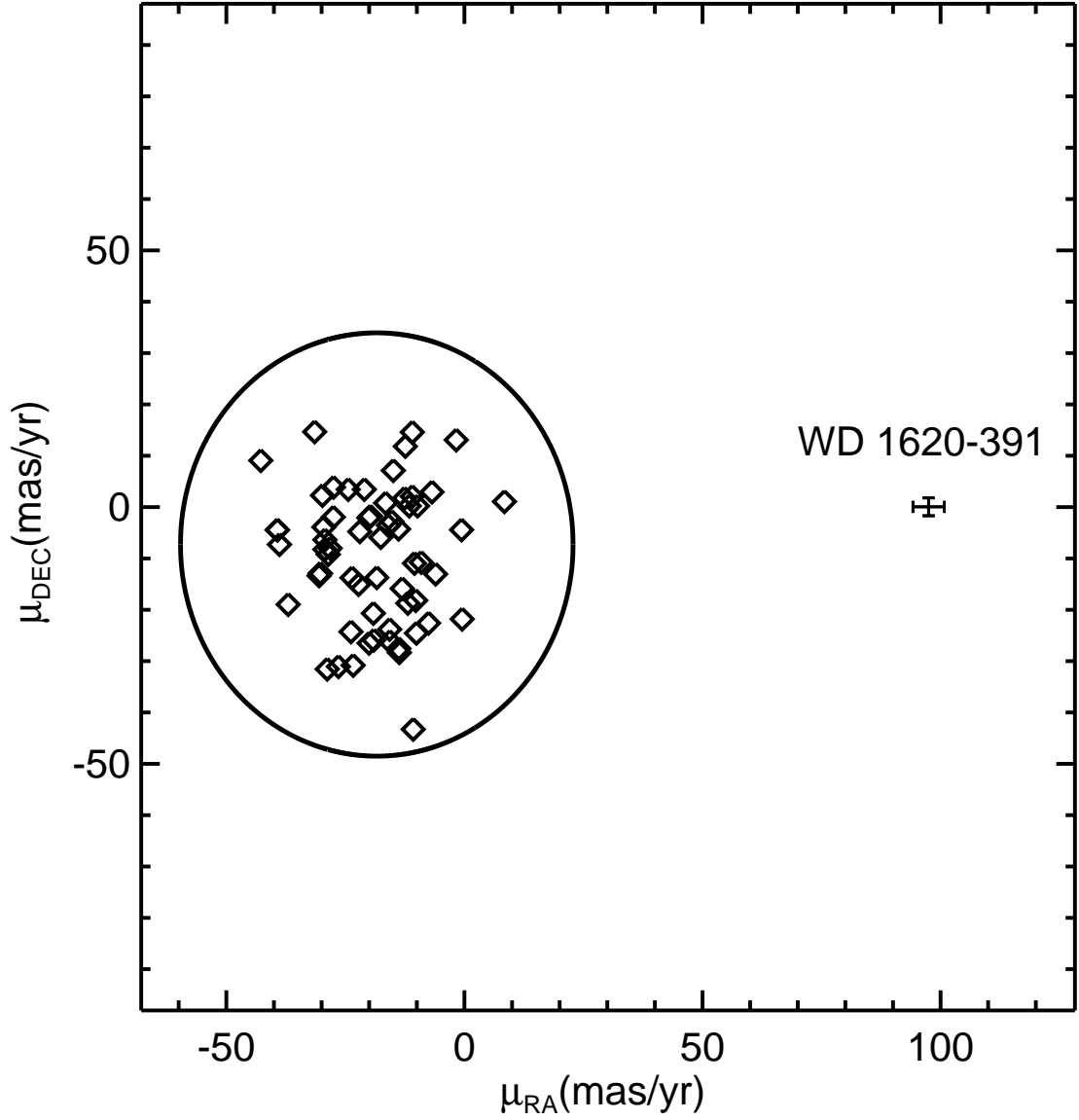}
\caption{\label{fig:wd1620prop} Measured proper motion in declination vs. proper motion in right ascension for observed sources close to WD 1620-391.  The
solid circle represents the 3$\sigma$ scatter of the observed objects, while
the square denotes where an object co-moving with WD 1620-391 would lie.}
\end{figure}

Figure \ref{fig:wd1620prop} shows the resulting differences between the measured 
centroids in the two NICMOS image epochs.  The solid circle represents the 3-$\sigma$ limit as empirically measured by the entire sample of observed sources
in the field, with 1$\sigma$ being 14 mas/yr and median proper motions
of the sample of -17 mas/yr and -8 mas/yr.  There is a slight offset in the
median change in right ascension of the group of sources from the expected zero value, though it is a $\sim$1$\sigma$ difference in RA.  This could be because of a bulk proper motion 
of the background sources, since WD 1620-391 is at a low galactic latitude,
or a sub pixel mismatch between the reported pixel centers of WD 1620-391.
The magnitude of centroiding errors on HST acquisitions, however, is closer
to 7~mas and is smaller than the offset seen here.  In any case, there 
appears to be no co-moving sources, thus completely ruling out any companions
down to 6M$_J$ at separations $>$ 13~AU \citep{debes05b}. 

\section{Conclusions}
\label{sec:conc}
 We can place stringent limits on the types of disks
and unresolved companions present for all of our targets.  For two of 
our targets, only planetary mass objects (M$<$14 M$_J$) can be present at all separations,
 and for the rest,
only very low mass brown dwarfs (M$<$ 20 M$_J$) 
can be present at separations $<$ 1\arcsec
or orbital separations of between 13 and 35~AU.

  The explanation that all apparently single DAZs can be caused by the winds of unseen companions
does not fit our results unless the companions are very low mass brown dwarfs
or high mass planets.
One would expect to see large amounts of dust present if tidally disrupted 
planetesimals or ISM accretion were the source of metals for DAZs.  
Our targets show no evidence of such dust down to $\sim$ 10$^{20}$ g if there
are optically thin disks present, and out to separations of $\sim$0.4 R$_\odot$
if there are optically thick disks present.  We effectively 
rule out optically thick disks like those seen around G~29-38 for our targets.
We cannot rule out dust that is further away from the white dwarf and consequently much cooler.  Sensitive studies at longer wavelengths may yet detect dust
around these white dwarfs.

Instead,
optically thick 
dusty disks around DAZs seem to be somewhat rare with only 5 such known and no
optically thin disks yet reported\citep{zuckerman87,
farihi05,kilic05,kilic06b,kilic07,farihi06}.  A lack of optically thick 
dust can be explained for cooler DAZs by infrequent encounters with 
large planetesimals that create short lived disks that disappear quickly
while still allowing detectable metal lines.  For that reason dusty 
disks should primarily be around hotter DAs, whose shorter settling times
require a quicker replenishment of dust and thus should have long lived
disks.  DAs that are too hot vaporize material well before it is tidally disrupted.  If the disks are instead optically thin, then weaker emission may be
present, though currently undetectable.  
The upper limits for dust disk masses imply that for many DAZs the amount
of material close to the white dwarf is sufficient to be detectable 
spectroscopically, but more difficult to detect in the mid-IR.

\acknowledgements
The authors would like to thank the anonymous referee for useful suggestions
in improving this paper.  We would like to thank Pierre Bergeron and Pier-Emmanuel Tremblay for helpful 
discussions on model white dwarf atmospheres and for graciously providing model
flux densities.
This work is based in part on observations made with the Spitzer Space Telescope, which is operated by the Jet Propulsion Laboratory, California Institute of Technology under a contract with NASA. Support for this work was provided by NASA through an award issued by JPL/Caltech.  Support for program \#10560 was provided by NASA through a grant from the Space Telescope Science Institute, which is operated by the Association of Universities for Research in Astronomy, Inc., under NASA contract NAS 5-26555.

%\bibliography{g29bib}
%\bibliographystyle{apj}

\clearpage

\begin{deluxetable}{llcccccccc}
%table of targets with age, mass, effective temperature
\tablecolumns{9}
\tablewidth{0pc}
\tablecaption{\label{tab:targs} Properties of the Target White Dwarfs}
\tablehead{
\colhead{WD} & \colhead{Name} & \colhead{M$_f$} &  \colhead{T$_{eff}$} & \colhead{t$_{cool}$} & \colhead{D} & \colhead{M$_i$\tablenotemark{a}} & \colhead{t$_{cool}$+t$_{MS}$} & $\tau_D$ & \colhead{References} \\
& & \colhead{(M$_\odot$)} & \colhead{(K)} & \colhead{(Gyr)} & \colhead{(pc)} & \colhead{(M$_\odot$)} & \colhead{(Gyr)} & log (yr) & }
\startdata
0208+396 & G 74-7 & 0.60 & 7310 & 1.4 & 17 & 2.1 & 3.2  &  3.78 & 1,4  \\
0243-026 & G 75-39 & 0.70 & 6820 & 2.3 & 21 & 3.2 & 2.8 & 3.39 & 1,4 \\
0245+541 & G 174-14 & 0.76 & 5280 & 6.9 & 10 & 4.6 & 7.2 & 4.47 & 1,4 \\
1257+278 & G 149-28 & 0.58 & 8540 & 0.9 & 34 & 1.7 & 3.3 & 3.26 & 1,4 \\
1620-391 & CD-38$^\circ$10980 & 0.66 & 24406 & 0.1 & 12 & 3.1 & 0.7 & & 2,3 \\
\enddata
\tablenotetext{a}{See Equation \ref{eqn:mass} for the calculation of 
M$_i$ and the WDs' total ages.}
\tablerefs{(1) \citet{bergeron01} (2) \citet{bragaglia95} (3) \citet{vanaltena95} (4) \citet{koester06}}
\end{deluxetable}

\begin{deluxetable}{cccccc}
\tablecolumns{6}
\tablewidth{0pc}
\tablecaption{\label{tab:obs} Observations}
\tablehead{
\colhead{WD} & \colhead{AOR Key} & \colhead{Exposure Time} & \colhead{Dither Points} & \colhead{Date} & \colhead{Start Time} \\
& & \colhead{(s)} & & & \colhead{(UT)}
}
\startdata
0208+396 & 11389184 & 100 & 30 & 2005-01-17 & 20:35:48 \\
0243-026 & 11389440 & 100 & 30 & 2005-01-16 & 15:44:34 \\
0245+541 & 11389696 & 100 & 30 & 2005-02-19 & 03:34:52 \\
1257+278 & 11389952 & 100 & 30 & 2005-06-13 & 03:18:19 \\
1620-391 & 11390208 & 30  & 75 & 2005-03-30 & 10:12:15 \\
\enddata
\end{deluxetable}

\begin{deluxetable}{ccccccccc}
\tablewidth{0pc}
\tablecolumns{9}
\tablecaption{\label{tab:fluxes}Predicted and Observed Fluxes in $\mu$Jy}
\tablehead{
\colhead{WD} & \colhead{[3.6]$_p$} & \colhead{[3.6]$_o$} & \colhead{[4.5]$_p$} & \colhead{[4.5]$_o$} & \colhead{[5.8]$_p$} & \colhead{[5.8]$_o$} & \colhead{[8.0]$_p$} & \colhead{[8.0]$_o$} \\
}
\startdata
0208+396 & 1039 & 1063$\pm$35 & 669 & 676$\pm$22 & 426 & 442$\pm$16 & 238 & 231$\pm$11 \\
0243-026 & 472 & 479$\pm$16 & 307 & 294$\pm$10 & 196 & 198$\pm$7 & 110 & 102$\pm$5 \\
0245+541 & 1333 & 1305$\pm$43 & 894 & 848$\pm$28 & 587 & 583$\pm$20 & 336 & 332$\pm$12 \\
1257+278 & 300 & 290$\pm$10 & 192 & 175$\pm$6 & 122 & 124$\pm$5 & 68 & 71$\pm$4 \\
1620-391 & 5100 & 5162$\pm$170 & 3204 & 3050$\pm$90 & 2006 & 2008$\pm$67 & 1097 & 1050$\pm$35\\
\enddata
\end{deluxetable}

\begin{deluxetable}{ccccc}
\tablewidth{0pc}
\tablecaption{\label{tab:limits} Excess Limits}
\tablehead{
\colhead{WD} & \colhead{Companion Limit} & \colhead{$i$\tablenotemark{a}} & \colhead{$R_{in}$\tablenotemark{b}} & \colhead{Dust Mass\tablenotemark{c}} \\
& \colhead{M$_J$} & \colhead{$i$} & \colhead{R$_\odot$} & \colhead{g}
}
\startdata
0208+396 & 20 & 2.9$^\circ$ & 0.7 & 2$\times$10$^{20}$ \\ 
0243-026 & 14 & 4.7$^\circ$ & 0.5 & 2$\times$10$^{20}$ \\
0245+541 & 20 & 1.6$^\circ$ & 0.4 & 1$\times$10$^{20}$ \\
1257+278 & 20 & 10.7$^\circ$ & 0.7 & 8$\times$10$^{20}$ \\
1620-391 & 13 & 0.1$^\circ$ & 5.0 & - \\
\enddata
\tablenotetext{a}{Upper inclination limit for optically thick disk to avoid
detection.}
\tablenotetext{b}{Lower limit for inner radius of optically thick disk.}
\tablenotetext{c}{Upper mass limit of dust for optically thin disk.}
\end{deluxetable}

\clearpage


\end{document}